\documentstyle[12pt,makeidx,bezier]{article}
\newcommand{\lb}[1]{\label{eqn:#1}}
\newcommand{\rf}[1]{\ref{eqn:#1}}

\begin{document}
\baselineskip 20pt

\centerline{\Large{\bf Symmetrization of Berezin Star Product}}
\centerline{\Large{\bf and Path-Integral Quantization}}

\vglue 1cm
\centerline{\large Satoru SAITO\footnote{email address: saito@phys.metro-u.ac.jp}\ \ and\ \ Kazunori WAKATSUKI\footnote{email address: waka@phys.metro-u.ac.jp}}
\vglue 1cm
\centerline{\it Department of Physics}
\centerline{\it Tokyo Metropolitan University}
\centerline{\it Minamiohsawa 1-1, Hachiohji}
\centerline{\it Tokyo, Japan 192-0397}

\vglue 2cm

{\abstract{
We propose a new star pruduct which interpolates the Berezin and Moyal quantization. A multiple of this product is shown to reduce to a path-integral quantization in the continuous time limit. In flat space the action becomes the one of free bosonic strings. Relation to Kontsevich prescription is also discussed.}}
 
\section{Introduction}

Since the discovery of Planck constant $\hbar$, the concept of quantization has been a central subject to study in physics during the last 100 years. The recent revival of the study of quantization in theoretical physics and in mathematics is aimed to quantize the base manifold, i.e., the space-time itself\cite{Yoneya}. From the view point of physics it must be founded in order to unify the theory of gravity into quantum (or gauge) field theory. It seems commonly understood that some new mathematics must be incorporated for the quantization of space-time to be established. There exist various methods of quantization which have been developed during 20th century. In order to understand the fundamental role of quantization in the new developement it will be useful to clarify relations among methods of quantization. 

It has been well known for some years that certain topological quantum field theories with Chern-Simons action can be considered within the framework of geometric quantization\cite{Verlinde}. The geometric quantization method\cite{Kirillov}\cite{Weinstein}\cite{Connes}\cite{Karasev} has been reformulated in terms of path-integral method\cite{AlekseevShatashivili}, from which conformal field theories are deduced. All these show that the geometric quantization is an important key word to understand integrable quantum field theory.

Besides the study of geometric quantization in field theory, there have been proposed direct methods of quantization. Among various deformation quantizations, the quantization of classical phase space using Moyal star product\cite{Moyal} is the one which has been well studied\cite{BFFLS},\cite{Sternheimer}, but is applicable only to flat phase space. Berezin quantization is another method which works in certain K\"ahler manifolds\cite{Berezin}\cite{Bar-Moshe}. The corresponding star product is given by involution of symbols associated with physical observables.

A constructive method of geometric quantization was proposed by Kontsevich\cite{Kontsevich}. It provides a systematic way to find coefficients of the expansion of the star product in powers of a deformation parameter. Subsequently this expansion was shown to be described in terms of path-integral of certain nonlinear sigma model\cite{Cattaneo Felder}\cite{Schomerus}. More recently Berezin quantization was generalized to work for an arbitrary K\"ahler manifold and an interpretation of the deformation was given in the language of Feynman diagrams\cite{Reshetkin-Takhtajan}.

Despite of many proposal of quantization methods, we are still far from understanding the real meaning of quantization from both physical and mathematical point of view. In particular relations between different methods are not obvious. For example all methods should be locally equivalent to the flat space quantization. The proof of this in the case of quantization of sphere a l\`a Berezin was given only recently\cite{Maeda}. The path-integral interpretation of the geometric quantization is quite interesting and has attracted many authors. It does not fill, however, the gap between the local nature of quantization and global nature of correlation functions, in spite of desire to provide a mathematical foundation of the path-integral quantization.
 
The aim of this paper is to clarify this point by introducing a new star product $f*g$. We adopt the spirit of Berezin quantization. At the same time we require that it turns to the Moyal bracket in a flat space, and reduces directly to a Poisson bracket in classical limit. We recall the fact that only the difference of the Berezin star product $(f*_Bg-g*_Bf)$ reproduces the Poisson bracket in the small parameter limit, while the Moyal star product $f*_Mg$ itself does. This difference made difficult to argue these two methods of quantization in equal footing for a long time. Therefore it is highly nontrivial to incorporate the two requirements simultaneously.

A multiple of this new product will be shown to reduce to a path-integral quantization in the continuous time limit. This explains the missing link between the naive concept of quantization and the intuitive construction of Feynman path-integral. 

Our star product does not show associativity in a manifest form, if it is not defined on a flat space. Nevertheless we can show that it plays the role of time ordered product, which is another important ingredient of field theory. The product becomes manifestly associative if we consider the path-integral representation of the product defined along a closed loop. In a flat space the action becomes one of free strings and associativity is satisfied as a result of duality of old string theory. The noncommutativity in string theory has its origin in the nature of Moyal star product\cite{Seiberg Witten}. Since our formalism of quantization is applicable to some K\"ahler manifolds, it generalizes the ordinary flat-space string theory. By the same reason we are able to make a correspondence of our method of quantization to the one of Kontsevich\cite{Kontsevich}.

\section{Moyal vs Berezin star products}

To begin with we consider the simplest case of Moyal star product:
\begin{eqnarray}
e^{i(m'x+n'p)}*_M\ e^{i(mx+np)}&:=&
\left.e^{i\lambda(\partial_{x'}\partial_{p}-\partial_{x}\partial_{p'})}e^{i(m'x'+n'p')}e^{i(mx+np)}\right|_{x'=x,p'=p}\nonumber\\
&=&e^{i\lambda(mn'-m'n)}e^{i(m'x+n'p)}e^{i(mx+np)}
\lb{Moyal star}
\end{eqnarray}
where $\lambda$ is a deformation parameter. Notice that the right hand side admits the following integral representation:
\begin{equation}
=
\int {d\xi' d\eta'\over 2\pi \lambda}\int {d\xi d\eta\over 2\pi \lambda}\ e^{i(\xi'\eta-\eta'\xi)/\lambda}e^{i[x(\eta'-\eta)-p(\xi'-\xi)]/\lambda}e^{i(m'\xi'+n'\eta')}e^{i(m\xi+n\eta)}.
\lb{Moyal product of exp}
\end{equation}
This information is sufficient to derive an integral representation of the Moyal star product of arbitrary functions which can be expanded into Fourier series. In fact we find the following expression:
\begin{equation}
(f*_M g)(x,p)=\int {d\xi' d\eta'\over 2\pi \lambda}\int {d\xi d\eta\over 2\pi \lambda}\ e^{i(\xi'\eta-\eta'\xi)/\lambda}e^{i[x(\eta'-\eta)-p(\xi'-\xi)]/\lambda}f(\xi',\eta')g(\xi,\eta).
\lb{Moyal product of func}
\end{equation}

In order to establish the correspondence between the Moyal star product and the Berezin form of quantization, let us rewrite above formulae in terms of complex variables defined by
$$
a=(x+ip)/\sqrt 2,\quad \bar a=(x-ip)/\sqrt 2,\quad b=(n-im)/\sqrt 2,\quad\bar b=(n+im)/\sqrt 2,
$$
\begin{equation}
v =(\xi+i\eta)/\sqrt 2,\quad \bar v =(\xi-i\eta)/\sqrt 2.
\lb{change of variables}
\end{equation}
Then $(\rf{Moyal product of func})$ is equivalent to
\begin{equation}
(f*_M g)(a,\bar a)
=\int{dv  d\bar v \over 2\pi i\lambda}\int{du  d\bar u \over 2\pi i\lambda}{\displaystyle{e^{-(v \bar v -v \bar a-a\bar v )/\lambda}}\over\displaystyle{e^{-(u \bar u -u \bar a-a\bar u )/\lambda}}}\ f(u ,\bar v )g(v ,\bar u ).
\lb{g*f}
\end{equation}
Here we denote by $f(a,\bar a)$ the same function as $f(x,p)$ but written in terms of the new variables. We also notice that the Moyal star product can be splitted into two parts:
\begin{equation}
(f*_{st} g)(a,\bar a):=e^{\lambda\partial_{a'}\partial_{\bar a}}f(a,\bar a)g(a',\bar a')
=\int{dv  d\bar v \over 2\pi i\lambda}e^{-(v \bar v -v \bar a-a\bar v +a\bar a)/\lambda}\ f(a,\bar v )g(v ,\bar a),
\lb{g*_stf}
\end{equation}
\begin{equation}
(f*_{ar} g)(a,\bar a):=e^{-\lambda\partial_{a}\partial_{\bar a'}}f(a,\bar a)g(a',\bar a')
=\int{d u  d\bar u \over 2\pi i\lambda}e^{( u \bar u - u \bar a-a\bar u +a\bar a)/\lambda}\ f( u ,\bar a)g(a,\bar u ).
\lb{g*_arf}
\end{equation}

We are now going to compare $(\rf{g*f})$ with Berezin star product. According to Berezin\cite{Berezin} we introduce a supercomplete set of state vectors $e^{K(a,\bar a')}$. Here $K(a,\bar a')$ is a function defined by an analytic continuation of the K\"ahler potential $K(a,\bar a)$ of the manifold in our concern. A multiplication law of covariant symbols is given by
\begin{equation}
(f*_B g)(a,\bar a):=\int d\mu(v,\bar v) e^{\Phi(v,\bar v,a,\bar a)/\lambda}f(a,\bar v)g(v,\bar a),
\lb{*_B}
\end{equation}
where $d\mu(v,\bar v)$ is the measure of integration and $\Phi$ is the Calabi function defined by
\begin{equation}
\Phi( v ,\bar v ,a,\bar a):=-K( v ,\bar  v )+K( v ,\bar a)+K(a,\bar v )-K(a,\bar a). 
\end{equation}
In a flat space the K\"ahler potential is $K(a,\bar a)=a\bar a$. The $*_B$ product $(\rf{*_B})$ does not reduce to the Moyal star product $(\rf{g*f})$ in the flat space, but does to the half star product $(\rf{g*_stf})$. In fact the Berezin star product was designed so that the $\lambda$ expansion of the commutator $(f*_B g\ -\ g*_B f)$ yields the Poisson bracket in the leading term, whereas the Moyal product $(\rf{Moyal product of func})$ contains the Poisson bracket in it. 

\section{New star product}

Now we would like to propose a new star product $*$, which fills the difference between the Moyal and Berezin star products :
\begin{equation}
(f*g)(a,\bar a):=\int d\mu( v ,\bar v ) \int d\mu( u ,\bar u ){\displaystyle{e^{\Phi( v ,\bar v ,a,\bar a)/\lambda}}\over\displaystyle{e^{\Phi( u ,\bar u ,a,\bar a)/\lambda}}}f( u ,\bar v )g( v ,\bar u ).
\end{equation}
Note that this single product $*$ reduces to the Moyal star product  $*_M$ of $(\rf{g*f})$ in the case of flat space. 

It is straightforward to see associativity of the Berezin star product $(\rf{*_B})$:
\begin{equation}
(f*_B g)*_B h=f*_B (g*_B h).
\end{equation}
In the case of our new product we have
\begin{eqnarray}
&&(f*\ (g* h))(a,\bar a)\nonumber\\
&=&
\int d\mu(v'',\bar\phi)\int d\mu(\phi,\bar v'')\int d\mu(v',\bar v)\int d\mu(v,\bar v')\ {\displaystyle{e^{\Phi(a,\bar a,\phi,\bar v'')/\lambda}}\over\displaystyle{e^{\Phi(a,\bar a,v'',\bar\phi)/\lambda}}}
{\displaystyle{e^{\Phi(\phi,\bar\phi,v,\bar v')/\lambda}}\over\displaystyle{e^{\Phi(\phi,\bar\phi,v',\bar v)/\lambda}}}\nonumber\\
&&\qquad\qquad\qquad\times f(v'',\bar v'')g(v',\bar v')h(v,\bar v)
\lb{f*(g*h)}
\end{eqnarray}
and
\begin{eqnarray}
&&((f*\ g)* h)(a,\bar a)\nonumber\\
&=&
\int d\mu(v',\bar v'')\int d\mu(v'',\bar v')\int d\mu(\phi,\bar v)\int d\mu(v,\bar \phi)\ {\displaystyle{e^{\Phi(v',\bar v'',\phi,\bar \phi)/\lambda}}\over\displaystyle{e^{\Phi(v'',\bar v',\phi,\bar\phi)/\lambda}}}
{\displaystyle{e^{\Phi(v,\bar\phi,a,\bar a)/\lambda}}\over\displaystyle{e^{\Phi(\phi,\bar v,a,\bar a)/\lambda}}}\nonumber\\
&&\qquad\qquad\qquad\times f(v'',\bar v'')g(v',\bar v')h(v,\bar v)
\lb{(f*g)*h}
\end{eqnarray}
which do not show associativity in an obvious manner. If the space is flat the integrations over $\phi$ and $\bar\phi$ yield Dirac $\delta$-functions. After some manipulation we find that
\begin{eqnarray}
&&(f*_M(g*_Mh))(a,\bar a)\nonumber\\
&=&
\int {d v'' d\bar v''\over 2\pi i \lambda}\int {d v' d\bar v'\over 2\pi i \lambda}\int {d v  d\bar v \over 2\pi i \lambda}\ f( v'',\bar v'')g( v',\bar v')h( v ,\bar v )\nonumber\\
&&\quad\times
\ e^{-( v\bar v' - v' \bar v)/\lambda}e^{a\bar v''-v''\bar a}\delta(a- v''+ v'- v )\delta(\bar a-\bar v''+\bar v'-\bar v ).
\end{eqnarray}
and
\begin{eqnarray}
&&((f*_Mg)*_Mh)(a,\bar a)\nonumber\\
&=&
\int {d v'' d\bar v''\over 2\pi i \lambda}\int {d v' d\bar v'\over 2\pi i \lambda}\int {d v  d\bar v \over 2\pi i \lambda}\ f( v'',\bar v'')g( v',\bar v')h( v ,\bar v )\nonumber\\
&&\quad\times
\ e^{-( v'\bar v'' - v'' \bar v')/\lambda}e^{v\bar a-a\bar v}\delta(a- v''+ v'- v )\delta(\bar a-\bar v''+\bar v'-\bar v ).
\end{eqnarray}
coincide with each other, hence the associativity is fulfilled. In case of general K\"ahler manifold the $\phi$ integration is not trivially done and the symmetry is not known.

\section{Path-integral via multiple star products}

Although we are not able to show the equivalence of $(\rf{f*(g*h)})$ and $(\rf{(f*g)*h})$ manifestly, we would like to show, in the following, that we can define properly the `ordered multiple products', which turn out to be state functions in path-integral representation in the continuous time limit.

Let us call the multiple product
\begin{equation}
\overleftarrow{A_{b\bar b}}^N(a,\bar a):=\left(e^{a\bar b^{(N+1)}-b^{(N+1)}\bar a} *\cdots *\ \left(e^{a\bar b^{(2)}-b^{(2)}\bar a} *\ \left(e^{a\bar b^{(1)}-b^{(1)}\bar a} *\ e^{a\bar b^{(0)}-b^{(0)}\bar a}\right)\right)\right),
\lb{forward multiple production}
\end{equation}
the `forward product'. Using the definition of the product repeatedly we obtain
\begin{eqnarray}
&=&
\int\prod_{j=1}^{N+1}d\mu(\phi^{(j-1)},\bar X^{(j)})d\mu(X^{(j)},\bar\phi^{(j-1)})
\left(\prod_{j=1}^{N+1}
{\displaystyle{e^{\Phi(\phi^{(j)},\bar\phi^{(j)},\phi^{(j-1)},\bar X^{(j)})/\lambda}}\over \displaystyle{e^{\Phi(\phi^{(j)},\bar\phi^{(j)},X^{(j)},\bar\phi^{(j-1)})/\lambda}}}\right)\nonumber\\
&&\qquad\times\ \prod_{j=0}^{N+1}e^{(X^{(j)}\bar b^{(j)}-b^{(j)}\bar X^{(j)})}
\end{eqnarray}
where we put $\phi^{(0)}:=X^{(0)},\ \phi^{(N+1)}:=a$.

Now suppose that the sequence of the product becomes infinitly large, so that $j$ turns to a continuous variable $\tau$. In this limit we have
\begin{equation}
\overleftarrow{A_{b,\bar b}}(a,\bar a)
=
\int{\cal D}\mu(\phi,\bar X){\cal D}\mu(X,\bar\phi)\ e^{iS/\lambda}
e^{\int^t_0  (X\bar b-b\bar X)d\tau},
\lb{continuous limit}
\end{equation}
where
\begin{equation}
S={1\over i}\int^t_0  d\tau 
\left(\pi{d\bar\phi\over d\tau}-{d\phi\over d\tau}\bar\pi\right),
\lb{action integral}
\end{equation}
and the canonical conjugate variables are defined by
\begin{equation}
\pi:={\partial K(\phi,\bar\phi)\over\partial\bar\phi}-{\partial K(X,\bar\phi)\over\partial\bar\phi},\qquad
\bar\pi:={\partial K(\phi,\bar\phi)\over\partial\phi}-{\partial K(\phi,\bar X)\over\partial\phi}.
\end{equation}
Notice that in the expression of $(\rf{continuous limit})$, we have put the boundary conditions $\phi(t)=a,\ \bar\phi(t)=\bar a$.

Similarly we define the `backward product'
\begin{equation}
\overrightarrow{A_{b\bar b}}^N(a,\bar a):=\left(\left(\left(e^{a\bar b^{(N+1)}-b^{(N+1)}\bar a} *\ e^{a\bar b^{(N)}-b^{(N)}\bar a}\right) *\ e^{a\bar b^{(N-1)}-b^{(N-1)}\bar a}\right) *\ \cdots *\ e^{a\bar b^{(0)}-b^{(0)}\bar a}\right).
\end{equation}
Using $(\rf{(f*g)*h})$ we obtain
\begin{eqnarray}
&=&
\int\prod_{j=0}^{N}d\mu(\phi^{(j+1)},\bar X^{(j)})d\mu(X^{(j)},\bar\phi^{(j+1)})
\left(\prod_{j=0}^{N}
{\displaystyle{e^{\Phi(X^{(j)},\bar\phi^{(j+1)},\phi^{(j)},\bar \phi^{(j)})/\lambda}}\over \displaystyle{e^{\Phi(\phi^{(j+1)},\bar X^{(j)},\phi^{(j)},\bar\phi^{(j)})/\lambda}}}\right)\nonumber\\
&&\qquad\times\ \prod_{j=0}^{N+1}e^{(X^{(j)}\bar b^{(j)}-b^{(j)}\bar X^{(j)})}
\end{eqnarray}
where $\phi^{(0)}=a,\ \phi^{(N+1)}=X^{(N+1)}$. In the continuous time limit this is exactly the same as the result of forward product $(\rf{continuous limit})$, except for the boundary conditions which we must impose $\phi(0)=a$ in the backward product, whereas we had put $\phi(t)=a$ in the forward product.  

The formulae we obtained provide us a natural interpretation of the products in terms of path-integration. Namely $S$ is nothing but a kinetic action of path-integral quantization. In the case of flat space the integration over $\phi$ and $\bar \phi$ can be performed in $(\rf{continuous limit})$, and we obtain a familiar path-integral representation of a free particle propagation:
\begin{equation}
\overleftarrow{A_{b,\bar b}}(a,\bar a)=\int{\cal D}X{\cal D}\bar X\ e^{iS/\lambda}e^{\int_0^t(X\bar b-b\bar X)d\tau},
\end{equation}
\begin{equation}
S={i\over 2}\int^t_0  d\tau X{d\bar X\over d\tau}
.
\end{equation}
\section{Associativity}

We notice that the asymmetry between the forward multiple product $\overleftarrow{A_{b,\bar b}}(a,\bar a)$ and the one of backward multiple product $\overrightarrow{A_{b,\bar b}}(a,\bar a)$  disappears if the path of integration over $\tau$ is closed to a loop and the condition $\phi(0)=\phi(t)=a$ is imposed. Knowing this information we are now going to provide a star product which manifestly satisfies associativity.

So far we have not discussed the external source functions $b^{(j)}$. We can define a generating functional by multiplying an arbitrary functional $w(b,\bar b)$ to $A_{b\bar b}(a,\bar a)$ and integrating over $b$'s:
\begin{equation}
F(w; a,\bar a)=\int{\cal D}b{\cal D}\bar b\ w(b,\bar b)\overleftarrow{A_{b,\bar b}}(a,\bar a).
\lb{F_w}
\end{equation}

As an example we consider the case
\begin{equation}
w(b,\bar b)=\delta\left(b(\tau)-\sum_{j=0}^{N+1}b^{(j)}\delta(\tau-\tau_j)\right)\delta\left(\bar b(\tau)-\sum_{j=0}^{N+1}\bar b^{(j)}\delta(\tau-\tau_j)\right),
\end{equation}
under the conditions $\phi^{(N+1)}=\phi(t)=a,\ \bar\phi^{(N+1)}=\bar\phi(t)=\bar a$. We also fix $b^{(0)}=\bar b^{(0)}=b^{(N+1)}=\bar b^{(N+1)}=0$. The substitution into $(\rf{F_w})$ yields
\begin{equation}
F(b^{(1)},\cdots,b^{(N)}; a,\bar a)=
\int_{a,\bar a}{\cal D}\mu(\phi,\bar X){\cal D}\mu(X,\bar\phi)\ e^{iS/\lambda}
\exp\left[\sum_{j=1}^N (X_j\bar b^{(j)}-b^{(j)}\bar X_j)\right],
\lb{X_j}
\end{equation}
where $X_j$ means the value of $X(\tau)$ at $\tau=\tau_j$.

Now suppose the path of integration over $\tau$ is closed, that is
\begin{equation}
S={1\over i}\oint d\tau 
\left(\pi{d\bar\phi\over d\tau}-{d\phi\over d\tau}\bar\pi\right)
\lb{closed action integral}
\end{equation}
in $(\rf{X_j})$. We further assume $\phi^{(0)}=\phi^{(N+1)}=a$ and $\bar\phi^{(0)}=\bar\phi^{(N+1)}=\bar a$. Then the result is manifestly symmetric under the cyclic permutation of $b^{(j)}$'s. It is more convenient if we multiply Fourier components $\tilde f_j$ of a function $f_j$ and integrate over $b^{(j)}$'s, from which we obtain
\def\ostar{*\kern -3mm {\rm o}}
\begin{eqnarray}
(f_N\ostar\cdots \ostar f_2\ostar f_1)(a,\bar a)&:=&\int\prod_{j=1}^Ndb^{(j)}d\bar b^{(j)}\ \prod_{j=1}^N\tilde f_j(b^{(j)},\bar b^{(j)})F(b^{(1)},\cdots ,b^{(N)}; a,\bar a)\nonumber\\
&=&
\int_{a,\bar a}{\cal D}\mu(\phi,\bar X){\cal D}\mu(X,\bar\phi)\ e^{iS/\lambda}\ \prod_{j=1}^Nf_N(X_j,\bar X_j).
\lb{ostar product}
\end{eqnarray}

The rule of the `ostar' product is as follows:
\begin{equation}
f_N\ostar\cdots \ostar\ f_{r+1}\ostar\ f_r\ostar\cdots\ostar\ f_1=(f_N\ostar\cdots \ostar\ f_{r+1})\ostar\ (f_r\ostar\cdots \ostar\ f_1)
\lb{factorizability}
\end{equation}
where the loops of integration are defined as\cite{Saito Kosterlitz}
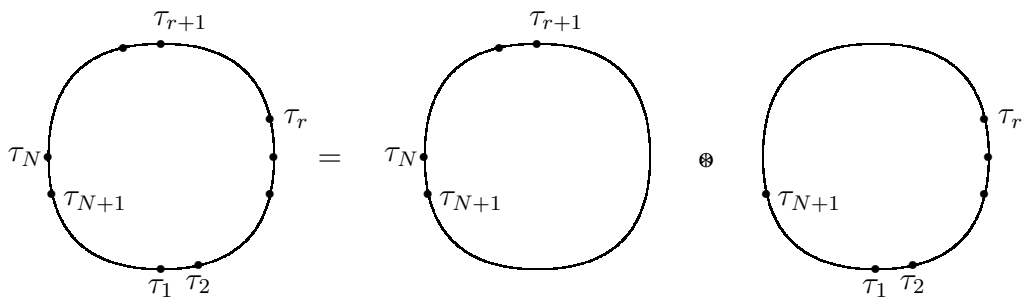
\begin{center}\begin{picture}(140,40)
\qbezier(5,20)(5,5)(20,5)\qbezier(5,20)(5,35)(20,35)\qbezier(20,35)(35,35)(35,20)\qbezier(35,20)(35,5)(20,5)
\put(20,5){\circle*{1}}
\put(19,1.5){\makebox(2,2)[c]{$\tau_1$}}
\put(25,5.5){\circle*{1}}
\put(24,2){\makebox(2,2)[c]{$\tau_2$}}
\put(34.5,15){\circle*{1}}
\put(35,20){\circle*{1}}
\put(34.5,25){\circle*{1}}
\put(37,24){\makebox(2,2)[c]{$\tau_r$}}
\put(20,35){\circle*{1}}
\put(19,37){\makebox(4,2)[l]{$\tau_{r+1}$}}
\put(15,34.5){\circle*{1}}
\put(5,20){\circle*{1}}
\put(1,19){\makebox(2,2)[c]{$\tau_N$}}
\put(5.5,15){\circle*{1}}
\put(7,13){\makebox(4,2)[l]{$\tau_{N+1}$}}
\put(42,19){\makebox(1,1)[c]{$=$}}
\qbezier(55,20)(55,5)(70,5)\qbezier(55,20)(55,35)(70,35)\qbezier(70,35)(85,35)(85,20)\qbezier(85,20)(85,5)(70,5)
\put(70,35){\circle*{1}}
\put(69,37){\makebox(4,2)[l]{$\tau_{r+1}$}}
\put(65,34.5){\circle*{1}}
\put(55,20){\circle*{1}}
\put(51,19){\makebox(2,2)[c]{$\tau_N$}}
\put(55.5,15){\circle*{1}}
\put(57,13){\makebox(4,2)[l]{$\tau_{N+1}$}}
\put(92,19){\makebox(1,1)[c]{$*$}}
\put(92,19){\makebox(1,1)[c]{o}}
\qbezier(100,20)(100,5)(115,5)\qbezier(100,20)(100,35)(115,35)\qbezier(115,35)(130,35)(130,20)\qbezier(130,20)(130,5)(115,5)
\put(115,5){\circle*{1}}
\put(114,1.5){\makebox(2,2)[c]{$\tau_1$}}
\put(120,5.5){\circle*{1}}
\put(119,2){\makebox(2,2)[c]{$\tau_2$}}
\put(129.5,15){\circle*{1}}
\put(130,20){\circle*{1}}
\put(129.5,25){\circle*{1}}
\put(132,24){\makebox(2,2)[c]{$\tau_r$}}
\put(100.5,15){\circle*{1}}
\put(102,13){\makebox(4,2)[l]{$\tau_{N+1}$}}
\end{picture}
\end{center}
on the left and right hand sides, respectively. Let us call this property of the ostar product `factorizability'. This is the property which quarantees associativity of the product:
\begin{equation}
f\ostar\ (g\ostar\ h)=(f\ostar\ g)\ostar\ h.
\end{equation}

Kontsevich had proposed a constructive method of star product which works for arbitrary Poisson manifold\cite{Kontsevich}. Subsequently it was shown being equivalent to a path-integral of a nonlinear sigma model\cite{Cattaneo Felder}. We obtain a corresponding formula from $(\rf{ostar product})$ by restricting $N=2$. Generalization of our formulation to Poisson manifold will be discussed elsewhere.

\section{Correspondence to string models}

All above discussions can be generalized to the case where several complex variables are present. Since a variable appears in a product with its canonical conjugate one, it amounts to reinterprete the product by their inner product. If there are infinitely many degrees of freedom, they can be combined into an inner product of fields. 

In order to illustrate the correspondence of our formulation to the string models discussed in early 70's\cite{Green Schwarz Witten}, we concentrate our consideration to the case of flat space. Under this circumstance the fields can be expanded into Fourier series:
\begin{equation}
 \bar X^\mu(\sigma)=x^\mu+\sum_{n=1}^\infty{1\over\sqrt n}\bar X^\mu_{n} e^{-in\sigma},\quad X^\mu(\sigma)=p^\mu\sigma-\sum_{n=1}^\infty{1\over\sqrt n}X^\mu_{n}e^{in\sigma}.
\end{equation}
The upper index $\mu=1,2,\cdots, d$ specifies the space time components. As we have discussed previously the integration over $X$ in the flat space can be easily done and the canonical conjugate of $X^\mu(\sigma)$ is nothing but $\bar X^\mu(\sigma)$ itself. Thus we find
\begin{equation}
\sum_{n=0}^\infty\left(X_{\mu n}{d{\bar X}^\mu_{n}\over d\tau}-{dX_{\mu n}\over d\tau}\bar X^\mu_{n}\right)
=
{1\over 2\pi i}\int_0^{2\pi}d\sigma\left({\partial X_\mu(\sigma)\over \partial\sigma}{\partial{\bar X}^\mu(\sigma)\over\partial\tau}-{\partial^2X_\mu(\sigma)\over\partial\tau \partial\sigma}\bar X^\mu(\sigma)\right),
\end{equation}
and the action is one of the free bosonic strings: 
\begin{equation}
S={1\over 4\pi}\int_0^t d\tau \int_0^{2\pi}d\sigma {\partial X_\mu\over\partial\sigma}{\partial\bar X^\mu\over\partial \tau }.
\lb{flat S}
\end{equation}

Let us consider a compact domain $D$ in the complex plane, whose boundary is given by $\sigma=0$. Instead of $(\tau,\ \sigma)$ we  use a complex coordinate $z$ to specify a point on $D$. Let $z_j,\ j=1,2,\cdots, N+1$ be $N+1$ points on the boundary which are fixed in this order corresponding to $\tau=\tau_1,\tau_2, \cdots, \tau_{N+1}$ and $k_1^\mu,k_2^\mu,\cdots, k_{N}^\mu$ be constant vectors associated with $\tau_1\cdots,\tau_N$. If we substitute
\begin{equation}
w(b,\bar b)=
C\oint \prod_{j=1}^{N+1}{dz_j\over z_j-z_{j+1}}
\delta\left(b^\mu(\tau)-\sum_{j=1}^{N}k_j^\mu\delta(\tau-\tau_j)\right)
\delta\left(\bar b^\mu(\tau)-\sum_{j=1}^{N}k_j^\mu\delta(\tau-\tau_j)\right)
\end{equation}
into $(\rf{F_w})$, we obtain
\begin{eqnarray}
&&F(k_1,k_2,\cdots, k_{N}; a,\bar a)\nonumber\\
&&=C\oint \prod_{j=1}^{N+1}{dz_j\over z_j-z_{j+1}}\int_{a,\bar a} {\cal D}X{\cal D}\bar X\ e^{iS/\lambda}
 \exp\left[i\sum_{\mu=1}^d\sum_{j=1}^{N}k_j^\mu\left(X_\mu(z_j)-\bar X_\mu(z_j)\right)\right],
\lb{string correlation fcn}
\end{eqnarray}
which represents $N$-point open string correlation function with external momenta $k_j^\mu$'s. In these formulae we have fixed three points, say $z_\alpha, z_\beta, z_\gamma$ among $z_j$'s, and denoted by $C$ the following quantity:
\begin{equation}
C^{-1}:={dz_\alpha dz_\beta dz_\gamma\over(z_\alpha-z_\beta)(z_\beta-z_\gamma)(z_\gamma-z_\alpha)},
\end{equation}
so that the integrations over $z_j$'s are not over counted. In particular we can choose $z_{N+1}=z_0$ being one of these three. Since we assumed $k_{N+1}^\mu=0$, the dependence on $X^\mu(z_{N+1})=X^\mu(z_0)=a^\mu$ is implicit. The three fixed points $(0,1,\infty)$ which appear in the Kontsevich formulation are to be identified with these points.

In this paper we have shown that the path-integral formalism of quantization arizes naturally starting from local concept of deformation quantization. The factorizability $(\rf{factorizability})$ and the cyclic permutation symmetry were the properties which lead to the string model of hadrons in 70's\cite{Kikkawa Sakita Virasoro}. This fundamental nature was called duality. The Moyal star product $(\rf{Moyal star})$ is equivalent to the product of vertices in operator formalism of strings. The `ostar' product, which generalizes the Moyal star product to non-flat manifolds, preserves this nature of duality. 

The string correlation functions have been well studied and known to appear as solutions to completely integrable systems, such as the Hirota bilinear identity\cite{Saito1} or the Yang-Baxter equation\cite{Saito2}. The duality has also relation to the pentagon equation\cite{Korepanov Saito}. Therefore it will be interesting to see the corresponding relations satisfied by the correlation functions defined using the new star product.
\vglue 1cm
\noindent
{\bf Acknowledgements}

The authors would like to thank Professors S.Iso, M.Karasev, I.Korepanov, Y.Maeda, N.Miyazaki, T.Tate and Mr.T.Masuda for useful discussions. This work is supported in part by the Grant-in-Aid for general Scientific Research from the Ministry of Education, Sciences, Sports and Culture, Japan (No 10640278).

\end{document}